\documentstyle[12pt,epsfig]{article}
\begin{document} 
\title{DNA-Protein Cooperative Binding through Long-Range Elastic Coupling.}
 
\author{Joseph Rudnick and Robijn Bruinsma \\ Department of Physics, 
UCLA \\ 405 Hilgard Ave., Los Angeles, Calfornia 90095-1547}

\date{\today} 
\maketitle
\begin{abstract}
Cooperativity plays an important role in the action of proteins bound 
to DNA.  A simple, mechanical mechanism for cooperativity, in the form of a 
tension-mediated interaction between proteins bound to DNA at two 
different locations, is proposed.  These proteins are {\em not} in 
direct physical contact. DNA segments intercalating bound proteins are 
modeled as a Worm-Like Chain, which is free to deform in two 
dimensions.  The tension-controlled protein-protein interaction is the 
consequence of two effects produced by the protein binding.  The first 
is the introduction of a bend in the host DNA and the second is the 
modification of the bending modulus of the DNA in the immediate 
vicinity of the bound protein.  The interaction between two bound 
proteins may be either attractive or repulsive, depending on their 
relative orientation on the DNA.  Applied tension controls both the 
strength and the range of protein-protein interactions in this model.  
Properties of the cooperative interaction are discussed, along with 
experimental implications.
\end{abstract}

\noindent {\small Keywords: protein, cooperativity, DNA, interactions}

\vspace{0.2in}

	The cooperative binding of proteins to DNA plays a significant role in 
the regulation of gene expression \cite{O-H} since it allows a 
sensitive response to small changes in protein concentration.  In 
particular, it is well known that transcription factor 
proteins \cite{Lodish} exhibit a significant level of 
cooperativity \cite{Sun}.  The structural basis of the cooperativity is 
not fully understood \cite{Sun}, but it is known that long-range 
cooperativity is possible through loops \cite{Schlief}, formed as the 
result of association between two DNA-binding proteins.  Looping is 
also believed to play an important role in gene access control.  
Cooperativity at shorter distances may be related to specific 
protein-protein interactions or to a generic cooperativity resulting 
from structural distortions induced by the binding of a protein to 
DNA \cite{Lilley,Nelson}.  For example, the binding of transcription 
regulation proteins such as the important TATA-box promoters (TPB) 
involves amino-acid intercalation into the stack of base pairs 
\cite{Kim,Werner}.  The result is that kinks are produced---sharp 
local bending angles in the DNA strand.  This deformation may permit a 
better fit for other DNA-associating proteins, such as the 
polymerases.  Disruptions of the base-pair stacking sequence have no 
effect beyond about half a turn of the double helix \cite{Kim,Werner}, 
so it is expected that this form of cooperativity is restricted to the 
immediate neighborhood of the primary binding protein.  
Protein-induced deformations of the DNA strand are not restricted to 
transcription factors.  DNA may {\em wrap} itself once or more around 
a protein as happens in the case of complexation of DNA with 
nucleosomes, the gyrase enzyme, or bacterial RNA.  Interestingly, 
nucleosome-binding appears to be cooperative with transcription 
factors \cite{O-H}.  DNA-deforming proteins will be referred to below 
as ``architectural'' proteins.

The aim of this article to demonstrate the possibility of a 
\underline{variable-range} form of cooperative DNA binding of architectural 
proteins with a range and strength that is regulated by the {\em tension} 
along the DNA strand.  The cooperative interaction between proteins 
that are {\em not} in physical contact, is mediated by the deformation of the 
intervening DNA strand.  In the absence of tension, the proposed 
mechanism is absent.  Our demonstration of the possibility of 
tension-controlled cooperativity is based on an analysis of the 
``Worm-Like Chain'' (WLC) model of DNA elasticity \cite{Hagerman,Kam}, 
which has been used in studies of single protein binding to DNA 
\cite{Marko}.  The WLC model is characterized by a single parameter, a 
length-scale $\xi_{p}$, known as the {\em persistence length}.  It is 
the distance over which a (tensionless) WLC maintains orientational 
order in the presence of thermal fluctuations.  That is to say, the 
autocorrelation between orientational order at two different 
locations of the chain falls off with distance, $\ell$, as 
$e^{-\ell/\xi_{p}}$.  Fitting the results of micro-mechanical {\em 
in-vitro} studies of protein-free DNA chains to the predictions of the 
WLC model yields good results for a persistence length of about 50 
nanometers \cite{Bustamante,Bensimon}, although longer persistence 
lengths have been reported by different methods \cite{Bednar}.

We employ the WLC model only to evaluate binding cooperativity due to 
deformations produced by architectural proteins in sections of the DNA 
strand that are not in the immediate neighborhood of the binding 
proteins themselves.  The WLC model will not apply reliably when the two 
proteins are so close together that details of base-pair action (i.e.  
roll, slide and twist) play an important role in the mediation of 
their interaction.  The primary binding of the protein itself is 
characterized by two phenomenological parameters: the single protein, 
zero-tension specific binding free energy $E^{(1)}(s)$ and the DNA 
bending angle $\alpha$; see Figure \ref{fig:twokinks}.  The specific 
binding energy $E^{(1)}(s)$ is a sequence-sensitive quantity that 
depends on the location of the protein along the chain.  It is usually 
in the range of 10 to 30 $k_{B}T$ with $k_{B}$ Boltzmann's constant 
and $T$ the temperature in degrees Kelvin.  At room temperature 
$k_{B}T \simeq 0.6$ kcal/mole.  By comparison, the individual ionic 
bonds between base pairs in DNA are in the range 2-3 kcal/mole, and 
typical covalent bounds are the order of 60-120 kcal/mole.  The index 
$s$ refers to the position of the protein along the DNA strand.  The 
protein-DNA complex will be assumed to be rigid for the tension levels 
envisioned, so the bending angle $\alpha$ is independent of tension.  
The values of $E^{(1)}(s)$ and $\alpha$ must be obtained either 
experimentally or by detailed structural modelling of DNA-protein 
interactions.

We will focus on the results of analytical and numerical studies that 
utilize the WLC model to compute the deformation energy of a DNA chain 
under tension with two identical architectural proteins attached and 
separated by a distance $l$.  The two proteins are assumed to induce a 
bend into the DNA strand without any twisting.  We find that there is, 
in general, both an \underline{enthalpic} and an \underline{entropic} 
contribution to the binding cooperativity proposed here \cite{Wang}.  
The enthalpic cooperative correction to the binding energy 
$E^{(2)}$ between the two proteins (denoted by 1 and 2) that are 
separated by a distance $l=|s_{1}-s_{2}|$ assumed large compared to 
the characteristic dimensions of the protein is given by
\begin{equation}
E^{(2)}_{\rm enth}\simeq E^{(1)}_{1}(s_{1}) + E^{(1)}_{2}(s_{2}) - 2 
\alpha^{2}k_{B}T \left(\frac{\xi_{p}}{\xi(F)}\right) \left(1 \mp 
e^{-|s_{1}-s_{2}|/ \xi(F)}\right)
\label{E2enth}
\end{equation}
This formula, derived in Appendix A, holds when the bending angle $\alpha$ 
is small compared to $\pi/2$.  The case of large bending angle is 
discussed later. The plus sign in Eq.  \ref{E2enth} refers to the 
``symmetric'' case with the two proteins bound on the same side of DNA 
while the minus sign refers to the ``anti-symmetric'' case with the 
two proteins bound on opposite sides of the DNA (see Figure 
\ref{fig:twokinks}).
The chain-tension $F$ in Eq.  \ref{E2enth} is variable but assumed to 
be in the the range of $10^{-2}$ to 10 pN.  The tension-dependent 
length-scale $\xi(F)$ in Eq.  \ref{E2enth} is of key importance; it 
sets the {\em range} of the binding cooperativity.  This quantity is defined 
by:
\begin{equation} 
\xi (F)=\sqrt {{{\xi _pk_BT} \over F}}
\label{xi(F)}
\end{equation}   
For a one pN tension, $\xi(F)$ is about 7 nm, while for a $10^{-2}$ pN force it 
is about 70 nm.  However, Eq.  \ref{E2enth} is only valid as long as the 
``tension-length'' $\xi(F)$ is less than the persistence length 
$\xi_{p}$.  For larger tensions $\xi(F)$ is small compared to the 
persistence length $\xi_{p} \approx 50 \ \mbox{nm}$.  Note that all 
parameters in Eq.  (\ref{E2enth}) can, in principle, be determined 
experimentally.

According to Eq.  \ref{E2enth}, there is no bending-induced 
cooperativity between two proteins separated by a distance large 
compared to the tension length $\xi (F)$.  In that limit, the only 
effect of tension is to reduce the single protein binding energy by an 
amount per protein $\Delta E_{\rm enth}=\alpha ^2\sqrt {k_BT\xi _pF}$ 
(using Eq.  \ref{xi(F)} to eliminate $\xi (F)$ in Eq.  
\ref{E2enth}).  This tension-induced reduction of the protein 
binding energy is discussed \cite{Marko} for the case of 
single-protein binding to DNA.  For a one pN tension, the binding 
energy reduction is significant: about $7 \alpha^{2}k_{B}T$.  With 
increasing tension, the protein will be released from the DNA chain 
when $\Delta E_{\rm enth}$ starts to approach the protein binding free 
energy $E^{(1)}$.  \footnote{ If the DNA-protein interaction 
involves a number of turns of the DNA around the protein, then we must 
add the quantity $\Delta LF$ to $\Delta E_{\rm enth}$, with $\Delta L$ 
the excess DNA length wound around the protein.  For low tensions $F$, 
this correction is small compared to $\Delta E_{\rm enth}$, but for a 
one pN tension, it is comparable.}

A summary of expected values for the cooperative interaction between 
two bound proteins is displayed in the table 1.  It is 
assumed that the bending angle that each enforces is $45^{\circ}$.  
Other quantities are appropriate to DNA at room temperature.

When the spacing between the two proteins is reduced to within a 
distance of order the tension-length $\xi (F)$, then the binding energy 
increases exponentially for the antisymmetric arrangement while it 
decreases exponentially for the symmetric arrangement.  We can 
interpret the last term in Eq.  \ref{E2enth} as an {\em effective 
potential energy of interaction}, $V_{\rm enth}\left( l \right)$, 
between two proteins given by:
\begin{equation} 
V_{\rm enth}(l)\cong \mp 2\alpha ^2k_BT\left( {{{\xi _p} \over {\xi 
(F)}}} \right) e^{-l/\xi (F)}
\label{Venth}
\end{equation} 
with $l=|s_{1}-s_{2}|$ the interprotein spacing. The minus sign in Eq.  
\ref{Venth} is for the anti-symmetric case.  In Appendix A we 
compute $V_{\rm enth}(l)$ for values of the bending angle $\alpha$ 
that range up to $\pi/2$.  The result is shown in Fig.  
\ref{fig:attractive}. Note that both the vertical and horizontal axes 
are dimensionless.  The energy has been expressed in units of 
$2k_{B}T\left(\xi_{p}/\xi(F) \right)$, the energy scale of the 
tension-induced binding energy reduction $\Delta E_{\rm enth}$ (see 
Eqs.  \ref{E2enth} and \ref{Venth}), while the distance is 
expressed in units of the tension length, $\xi(F)$.  It follows from 
Fig.  \ref{fig:attractive} that an increase in tension {\em reduces} 
the range of the interaction, as expected from Eq.  \ref{xi(F)}, 
while it {\em increases} the strength of the cooperativity.  If the 
spacing $l$ between the proteins is small compared to the tension 
length $\xi(F)$, then the effective potential $V(0)$ cancels the 
$\Delta E_{\rm enth}$ term in Eq.  \ref{E2enth}.  The enthalpic 
energy gain obtained by bringing two proteins together along the chain 
from a large separation is equal to twice $\Delta E_{\rm enth}$.

	For the symmetric case, the effective potential energy is repulsive. The 
corresponding energy plots are shown in Figure \ref{fig:repulsive}.
The enthalpic deformational energy now  increases as the two 
proteins approach each other.  In the limit of small bending angle 
$\alpha$, two adjacent bending proteins with $\alpha$ in the symmetric 
conformation have the same tension-induced binding energy reduction as 
a single protein with a double bending angle of $2 \alpha$.  The 
energy scale for the cooperativity is thus in general set by $\Delta 
E_{\rm enth}$.  The effective interaction potential is, then, always 
less than the single protein binding energy since proteins are 
expected to unbind when $\Delta E_{\rm enth}$ spacing approaches 
$E^{(1)}(s)$.

Although it would appear as if the enthalpic cooperativity depends on 
temperature within the WLC model (see Eqs.  \ref{E2enth} and \ref{Venth}), 
this is not the case: $\xi_{P}$ is inversely proportional to $k_{B}T$ 
(see Appendix A) so neither $\Delta E_{\rm enth}$ nor $\xi(F)$ depend 
on $k_{B}T$.  There {\em is}, however, a purely entropic contribution 
to the cooperativity that is explicitly dependent on $k_{B}T$.  In 
Appendix B, we obtain the following expression for the entropic correction 
to the cooperativity:
\begin{equation}
\Delta E_{\rm ent}^{(2)}=k_BT\left\{ {{d \over {\xi (F)}}- \ln \left( 
{1+{d \over {\xi (F)}}+{1 \over 4}\left( {{d \over {\xi (F)}}} 
\right)^2\left[ {1-e^{-{{2l} \over {\xi (F)}}}} \right]} \right)} 
\right\}
\label{Eentr1}
\end{equation}
for two proteins of length $d$ separated by a distance 
$l=|s_{1}-s_{2}|$.  The entropic contribution does not depend on the 
bending angle $\alpha$ and is the same for the symmetric and 
antisymmetric configurations.  In the limit of protein separations that 
are large compared to the tension length $\xi(F)$, the entropic 
contribution $\Delta E_{\rm entr}={1 \over 2}\Delta E_{\rm 
entr}^{(2)}\left( \infty \right)$ to the single-protein binding energy 
is:
\begin{equation}
\Delta E_{\em entr}=k_BT\left\{ {{d \over {2\xi (F)}}- \ln \left( {1+{d 
\over {2\xi (F)}}} \right)} \right\}
\label{Eentr2}
\end{equation}
This is, again, a negative quantity: the local constraints imposed on the 
DNA chain by the two binding proteins lowers the entropy of the complex 
as compared to a free chain and hence reduces the binding energy.  The 
difference can, again, be interpreted as an effective entropic potential 
energy of interaction, which is now entropic.  It is given by
\begin{equation}
V_{\rm entr}(l)=k_BT\ln \left( {1-{{\left( {{d \over {\xi (F)}}}
\right)^2e^{-{{2l} \over {\xi (F)}}}} \over {4\left( {1+{d \over {2\xi (F)}}
} \right)^2}}} \right)
\label{Ventr} 
\end{equation} 
This effective entropic potential energy is always 
attractive.  In Fig.  \ref{fig:Vgraph} we show the entropic potential 
energy for two proteins of size d equal to 20 \AA ngstroms and for a 
one pN applied tension ($\xi(F=1 \mbox{pN})=70$ \AA).

For larger bending angles, the entropic interaction is both weaker and 
shorter in range than the enthalpic interaction.  However, since this 
interaction does not depend on the magnitude of the bending angle, it 
dominates for zero bending angles, or bending angles that are very small.  An 
interesting special case concerns the entropic interaction between two 
long \underline{strings} of binding proteins.  If we model a 
polymerized string of proteins bound to DNA as a rigid section of size 
$d$, with $d$ assumed large compared to the tension length, and with zero 
total bending angle, then two such strings are expected to have an 
effective entropic interaction potential given 
by:\footnote{Formally, there is a divergence in Eq.  \ref{Vlim} for 
small spacings $l$, but Eq.  \ref{Vlim} should, of course, not be 
expected to retain validity when $l$ approaches a base-pair spacing.}
\begin{eqnarray} 
V_{\rm entr}(l) &\cong& k_BT\ln \left( {1-e^{-{{2l} \over {\xi (F)}}}} 
\right) \nonumber \\
d/\xi (F)&\to& \infty
\label{Vlim}
\end{eqnarray}
The free energy of two long rigid strings separated by a small gap is 
lowered by an amount of order $k_{B}T$ if the intervening gap is 
filled in either by shifting one of the two strings to close the gap 
or by adding additional binding proteins inside the gap.  This 
effect provide us with a curious {\em entropic} stabilization of 
mechanism of polymerization of proteins along DNA strands.

When we increase the bending angle beyond $\pi/2$, new physical 
effects appear.  As shown in Figures \ref{fig:loop} and 
\ref{fig:repulsivepair}, there are in general two possible configurations 
for a symmetric two protein/DNA complex.  Up to now, it has been tacitly 
assumed that the ``S'' or ``stretched'' configuration was the appropriate one 
(as shown in Fig.\ref{fig:twokinks}) and indeed the S configuration 
has the lower free energy for bending angles $\alpha$ less than $\pi/2$.  
However, when the bending angle exceeds $\pi/2$ this is no longer the 
case.  For low tensions, the ``L'', or ``looped'' configuration has in fact 
the lower elastic free energy while for higher tensions, the S 
configuration is more stable.  The two regimes are separated by a 
mathematical singularity which has the character of a 
\underline{first-order phase transition}.  In Fig.  
\ref{fig:transition} we show the extension, $X$, of the two-kink 
configuration for which the kink angle is greater than $\pi/2$, as a 
function of tension,$F$.  There is a transition from the L to the S 
configuration with increasing tension visible as a discontinuity of 
the extension at the transition point.  There is no transition in the 
antisymmetric case.

	In summary, we have shown that, within the confines of the WLC model, 
tension can trigger cooperative binding for anti-symmetrically arranged 
architectural proteins.  The binding strength has both an enthalpic 
and an entropic contribution, and it has an appreciable magnitude for 
tensions of the order of one pN or higher.  For larger bending angles, 
we find two competing configurations connected by a tension-induced 
phase-transition.  Both the enthalpic and entropic contributions to 
cooperativity vanish in the limit of zero tension (see Eqs.  
\ref{E2enth} and \ref{Ventr}).  This last result is certainly not 
self-evident.  The decrease in chain entropy imposed by two rigid 
sections can, for instance, be expected to depend on the spacing, even 
for zero tension.  Interestingly, a study of the interaction between 
two stiff inclusions inside a {\em two dimensional} surface (such as a 
membrane) reports \cite{Bruinsma} that in this case there is a 
long-range zero-tension interaction with both entropic and enthalpic 
contributions, both dropping off as the inverse fourth power of the 
spacing between the inclusions.  The disappearance of tension-induced 
cooperativity at $F=0$ thus must be related to the fact that we are 
dealing with a one-dimensional geometry.

Experimental {\em in vitro} tests of the proposed mechanism can be 
performed by preparing a bundle of DNA strands, each strand containing 
bacterial gene operator sequences periodically spaced by a distance of 
$l$ base pairs.  The associated repressor protein (such as the lac 
repressor) binding specifically to the operator sites would induce 
local kinks at the operator sites.  According to the model, the 
logarithm of the equilibrium repressor-operator binding constant 
$K_{\rm RO}$ contains a contribution that depends on the operator spacing 
$l$ and the tension $F$ of the DNA bundle according to Eq.  
\ref{E2enth}.

A study of the \underline{kinetics} of cooperative protein-DNA 
association can also be a testing ground.  For instance, since the TATA box 
binding protein TBP is known to produce a large bending angle, the 
one-dimensional diffusion along the DNA of other proteins required for 
the RNA polymerase initiation complex, such as TFIIE, H and J, that 
are non-specifically bound to DNA will be speeded up by 
bending-induced cooperativity.  The reason is that the effective 
potential $\Delta V(l)$ turns the random one-dimensional diffusion 
into a \underline{directed} process.  It should be noted here that the 
weaker non-specific binding of DNA associating proteins will not 
deform the DNA strand as much as specific binding.  However, studies 
of the dependence of non-specifically bound 434 repressors on the DNA 
flexibility indicate that non-specific bonding also involves 
distortions of the local DNA structure, so there still ought to be a 
tension-controlled interaction between specific and non-specifically 
bound proteins \cite{Hogan}.  We thus predict that (modest) tension 
will actually \underline{increase} the formation rate of the RNA 
polymerase initiation complex.  At high tension levels, the formation 
rate will decrease with tension for reasons discussed earlier.

Another possible area where the present theory could be applied is 
histone-DNA interactions.  According to our model calculations, a 
collection of non-specifically bound proteins ought to adopt an 
antisymmetric zig-zag configuration under tension.  The binding of 
histones to DNA is reported to produce a zig-zag nucleosome structure 
consistent with our calculations \cite{Thoma}.  Under tension, the 
zig-zag structure should thus be stabilized. It should be kept in mind, though, 
that the intervening linker histones may well affect the competition 
between different configurations \cite{Thoma}.

An important question for the relevance of the work presented here is 
whether DNA is under tension under {\em in-vivo} conditions.  DNA strands in 
suspended in good solvent are in fact not under tension.  We believe 
that this is an exceptional case.  A DNA strand whose ends are fixed 
is, even for low extensions, typically under a tension in the range of 
0.01 pN due to thermal fluctuations, as shown by the micromechanical 
studies \cite{Bustamante,Bensimon}.  For extensions closer to one, the 
tension can be much larger.  If a DNA strand is subject to the 
activity of force-transducing proteins---for instance during mitosis, 
RNA transcription, or homologous recombination---then much higher 
tensions can be generated.  It is known that a single motor protein is 
able to generate a force of 10 pN or more \cite{Yin}.  Finally, 
architectural proteins themselves generate tension if they attach to a 
DNA strand with fixed ends.  When an architectural protein attaches, 
DNA material is required to accomodate the deformation of the DNA 
chain near the protein.  A simple calculation shows that the 
self-induced tension of a DNA strand of length $L$ whose ends are held 
fixed a distance $X$ apart and which contains a line density $\rho$ of 
bending proteins is given by:
\begin{equation}
F\cong {{4k_BT\left\langle {\alpha ^2} \right\rangle \xi _p} 
\over {\left( {1-{X \over L}} \right)^2}}\rho ^2
\label{sit}
\end{equation}
with $\langle \alpha^{2}\rangle$ the average of the square of the bending 
angle.  If the line density is of the order of one protein per hundred 
Angstrom and if $\langle \alpha^{2}\rangle$ is of the order one, then 
this self-generated tension 
is of the order one pN (it should be possible to verify Eq.  \ref{sit} in 
micro-mechanical measurements). It thus seems reasonable to assume that DNA is 
under tension for {\em in-vivo} conditions.

Our results were derived with DNA-protein binding in mind, but the 
general aspects of our conclusions are related to work in other areas.  
We can consider the predicted aggregation of proteins under tension as 
a form of stress-induced \underline{decomposition}.  Stress-induced 
phase-separation of multi-component systems is actually a classic 
phenomenon in solid-state materials for the case that different 
constituents have different elastic moduli \cite{Cahn}, just as 
envisioned in the present case.

We conclude by noting that a number of technical objections can be raised 
against the method used in our study.  It is not really reasonable to 
assume that DNA twist plays no role, in view of the helical nature of 
DNA, and that we can be allowed to restrict ourselves to a purely 
two-dimensional arrangement.  However, it is our current belief that 
the introduction of the twist degree of freedom introduces 
qualitative, but not quantitative, modifications to the results 
reported here.  The assumption of internal structural rigidity of 
proteins also is questionable.  For instance, it is well known that 
many enzymes can undergo stress-induced structural changes.  Enzymes 
bound to DNA also may change their structure under stress.  Finally, 
it must be kept in mind that the WLC model neglects the possible 
influences of intervening structures, such as other bound proteins, on the 
tension-induced interaction betwen two bound proteins, and that 
looping and other manifestations of DNA self-interaction are ignored.  
Despite all these caveats, we feel that the basic 
result---stress-induced aggregation of DNA-bound proteins---is robust 
and should remain present in more realistic models.

\section*{Acknowledgements} 
We would like to thank C. Bustamante, 
D.  Chatenay, and E.  Siggia for helpful conversations.  We are 
especially grateful to P.  O'Lague for useful comments. R.  B.  acknowledges 
the support of the NSF through grant DMR-9708646.

\begin{appendix}
\renewcommand{\theequation}{A.\arabic{equation}}
\setcounter{equation}{0}
\section*{Appendix A}
We model the DNA strand as a rod that is free to move in two dimensions. 
The shape of the rod is parameterized in terms of an angular variable 
$\theta$, which will vary with distance, $s$, along the rod.  This 
parameterization is pictured in Figure \ref{fig:parameters}.  The 
energy of a configuration of the system is given by the following 
expression.
\begin{equation} 
H\left[ {\theta \left( s \right)} 
\right]=\int{ds\left\{ {{K \over 2}\left( {{{d\theta } \over {ds}}} 
\right)^2-F\cos \theta } \right\}}
\label{energy}
\end{equation}
The rigidity against bending is parameterized in terms of a stiffness 
parameter $K$, while $F$ is the tension applied to the rod.  The 
quantities referred to in the text are related to $K$ and $F$ by $\xi 
\left( F \right)=\sqrt {K/F}$ and 
$\xi _p=K/k_BT$.  In 
the ``classical'' limit, which applies at very low temperatures, is 
determined by the extremum equation
\begin{equation}
{{\delta H} \over {\delta \theta \left( s \right)}}=
-K{{d ^2\theta } \over {d s^2}}+F\sin \theta \left( s \right)=0
\label{extremum}
\end{equation} 
This equation is solved by quadratures by noting that it is 
identical to the first order differential equation
\begin{equation}
{K \over 2}\left( {{{d\theta \left( s \right)} \over {ds}}} 
\right)^2+F\cos \theta \left( s \right)=\kappa 
\label{quads1}
\end{equation} 
with $\kappa$ a constant.  From Eq.  \ref{quads1} we immediately obtain
\begin{equation}
{{d\theta } \over {\sqrt {{{2F} \over K}
\left( {\Lambda -\cos \theta } \right)}}}=\pm ds
\label{quads2} 
\end{equation}
with $\Lambda$ another constant.  Integration of \ref{quads2} yields an 
implicit form for $\theta(s)$, in which $s$ is represented as a 
function of $\theta$.  The specific function is a combination of 
elliptic integrals.  There is no evident way to invert the expression 
thus derived to obtain $\theta$ as an explicit function of $s$.

On the other hand, when the bending of the DNA is small, so that $\theta$ 
is small, the cosine function in \ref{energy} can be expanded.  The zeroth 
order term is a constant, ``background'' energy.  The energy of the 
system, as it depends on the configuration of the DNA, is given by
\begin{equation}
H\left[ {\theta \left( s \right)} \right]=\int {ds\left\{ {{K \over 
2}\left( {{{d\theta \left( s \right)} \over {ds}}} \right)^2+{F \over 
2}\theta \left( s \right)^2} \right\}}
\label{smallangleE}
\end{equation} 
The statistical mechanics of the system controlled by 
this energy can be analyzed in complete detail.  Here, the extremum 
equation is
\begin{equation}
-K{{d^2\theta \left( s \right)} \over {ds^2}}+F\theta \left( s \right)=0
\label{saextreme}
\end{equation} 
the solution of which is
\begin{equation}
\theta \left( s \right)=Ae^{s/\xi \left( F \right)}+
Be^{-s/\xi \left( F \right)}
\label{sasol}
\end{equation}
Where
$\xi \left( F \right)=\sqrt {KF}\equiv \sqrt {\xi _pk_BTF}$, and $A$  and $B$ 
are constants.

As an example of the use of the small angle formulas 
\ref{smallangleE}-\ref{sasol} we calculate the free energy cost of the 
presence of two proteins that kink the DNA to which they are attached.  
These insertions enforce a discontinuity in $d \theta / ds$ at the 
location of each kink.  We assume that the magnitude of the 
discontinuity is $\alpha_{-}$ at the first kink and $\alpha_{+}$ at the second.  
To simplify the analysis, we place the first kink at $s=-l/2$ and the 
second kink at $s=l/2$.  The calculation of the classical solution 
subject to these constraints is relatively straightforward.  One finds
\begin{equation} 
\theta \left( s \right)=\left\{ \matrix{\theta _<e^{\sqrt {{F \over 
K}}\left( {s+{l \over 2}} \right)}&&&&&&s<-{l \over 2}\hfill\cr 
-\theta _-{{\sinh \sqrt {{F \over K}}\left( {s-{l \over 2}} \right)} 
\over {\sinh \sqrt {{F \over K}}l}}+\theta _+{{\sinh \sqrt {{F \over 
K}\left( {s+{l \over 2}} \right)}} \over {\sinh \sqrt {{F \over 
K}l}}}&&&&&&-{l \over 2}<s<{l \over 2}\hfill\cr \theta _>e^{-\sqrt {{F 
\over K}}\left( {s-{l \over 2}} \right)} &&&&&&{l \over 2}<s\hfill\cr} 
\right.
\label{aandb}
\end{equation}
According to \ref{aandb}, the angle is equal to $\theta_{<}$ immediately to 
the right of $s=-l/2$ and $\theta_{-}$ immediately to the left of 
$s=-l/2$.  By the same token $\theta(l/2-\epsilon)=\theta_{+}$ and 
$\theta(l/2+\epsilon)=\theta_{>}$.  The kinks give rise to a 
discontinuity at $s=\pm l/2$, which means that $\theta_{<} \neq 
\theta_{-}$ and $\theta_{>} \neq \theta_{+}$.  We write
\begin{eqnarray}
\theta _<&=&\theta _--\alpha _-\cr \nonumber \\ 
\theta _>&=&\theta _++\alpha _+
\label{thetadefs}
\end{eqnarray}
Then, the total energy, $E$, of the bent rod, as given by Eqs.  
\ref{smallangleE}, \ref{aandb} and \ref{thetadefs}, is given by 
\begin{equation}
E={1 \over 2}\sqrt {FK}\left\{ {\left( {\theta _--\alpha _-} \right)^2+
\left( {\theta _++\alpha _+} \right)^2+\theta _+^2\coth \sqrt {{F \over K}}l+
\theta _-^2\coth \sqrt {{F \over K}}l+{{2\theta _+\theta _-} 
\over {\sinh \sqrt {{F \over K}}l}}} \right\}
\label{Eresult1}
\end{equation}
If we replace $\theta_{+}$ by $\Sigma + \Delta$ and $\theta_{-}$ by $ 
\Sigma - \Delta$, then Eq.  \ref{Eresult1} becomes
\begin{eqnarray} 
E\left( {\Sigma ,\Delta } \right)= \nonumber \\
&\sqrt {FK}\left\{ {\Sigma ^2{{e^{\sqrt {{F \over K}}l}+1} \over 
{\sinh \sqrt {{F \over K}}l}}+\Delta ^2{{e^{\sqrt {{F \over K}}l}-1} 
\over {\sinh \sqrt {{F \over K}}l}}+\Sigma \left( {\alpha _+-\alpha 
_-} \right)+\Delta \left( {\alpha _++\alpha _-} \right)+\alpha 
_+^2+\alpha _-^2} \right\} \nonumber \\
\label{Eresult2} 
\end{eqnarray} 
The partition function, $Z$, of the system 
is given by
\begin{equation}
Z = \int \int d \Sigma d \Delta \exp \left( -\beta E \left( \Sigma, 
\Delta \right) \right)
\label{Zdef}
\end{equation}
where $\beta = 1/k_{B}T$.  This double gaussian integral is evaluated by 
completing squares.  Taking the log and multiplying by 
$k_{B}T$ to obtain the free energy, $F$, we find 
\begin{equation}
F={{\sqrt {FK}} \over 4}\left\{ {\alpha _+^2+\alpha _-^2+2\alpha _+\alpha _
-e^{-\sqrt {{F \over K}}l}} \right\}-{{k_BT} \over 2}\ln \left( {1-e^{-2\sqrt 
{F \over K}l}} \right)
\label{Fresult1} 
\end{equation}
The first term on the right hand side of the expression for the 
free energy in Eq.  \ref{Fresult1} is the mean field approximation to the free 
energy cost of two kinks in the rod.  The second term is a partial 
contribution to the free energy due to fluctuations, in particular 
fluctuations in the angles $\theta_{+}$ and $\theta_{-}$.  To complete the 
evaluation of the free energy associated with a pair of kinks, we must 
now average over all other fluctuations in the rod.  To do this, we 
expand $\theta (s)$ in normal sinousoidal modes, subject to the condition that 
those modes do not alter the values of $\theta$ immediately to the right or 
the left of the locations of the kinks.  The contribution to the free 
energy that depends on the distance, $l$, between the kinks arises from 
the fluctuations in that region.  This fluctuation sum reduces to 
\begin{equation}
k_BT\left\{ {\sum\limits_{n=1}^\infty  {\ln \left[ {\left( {{{n\pi } \over l}} 
\right)^2+{F \over K}} \right]-{1 \over 2}\ln {F \over K}}} \right\}={{k_BT} 
\over 2}\ln \left( {1-e^{-\sqrt {{F \over K}}l}} \right)
\label{fluctuation}
\end{equation}
In arriving at the expression on the right hand side of Eq.  
\ref{fluctuation}, an infinite contribution to the sum over logarithms was 
subtracted off.  This ``regularization'' of the sum is consistent with 
standard field theoretical approaches to the evaluation of the free 
energy of systems such as the one under consideration here.  The right 
hand side of \ref{fluctuation} \underline{exactly cancels} the fluctuation 
contribution to \ref{Fresult1}.  Thus, the free energy of the two-kink system 
consists entirely of the mean-field contribution.

\section*{Appendix B}
\renewcommand{\theequation}{B.\arabic{equation}}
\setcounter{equation}{0}

In the limit that the kinks occupy an infinitesimal portion of the DNA, 
the free energy cost of a pair of them is, as noted in the appendix 
above, given entirely by mean-field theory.  There is a mechanism 
leading to fluctuation induced interaction energy, and that is the 
modification of the bending modulus by the proteins that cause the 
kinks.  This modification can be modeled as follows.  In the 
small-angle approximation, one replaces the energy expression in 
\ref{smallangleE} by
\begin{equation}
H\left[ {\theta \left( s \right)} \right]=
\int {ds\left\{ {{{K_{1,2}} \over 2}\left( {{{d\theta \left( s \right)} 
\over {ds}}} \right)^2+{F \over 2}\theta \left( s \right)^2} \right\}}
\label{variousK}
\end{equation} 
where the symbol $K_{1,2}$ stands for the two possible values, $K_{1}$ 
and $K_{2}$, that the bending modulus can now take on in the various 
regions.  The subscript 2 applies in the regions that are in the 
immediate vicinity of the proteins, while the subscript 1 is 
appropriate everywhere else.  This means that $K_{1}$ is the 
unsubscripted $K$ in previous expressions.  The effects of 
modifications of the bending modulus can be calculated separately from 
the consequences of the bending of the DNA by proteins.  This means 
that we can ignore any discontinuities or alterations of the 
equilibrium value angle $\theta(s)$ that result from the presence of 
proteins.

The regions in which the bending modulus takes on its two possible 
values are indicated in Figure \ref{fig:parameters}. The bending modulus is 
equal to $K_{2}$ in the two heavily drawn regions, while it is equal 
to $K_{1}$ everywhere else.  Matching conditions at the boundary 
between regions are that $\theta(s)$ and $K_{i} d\theta/ds$ are 
continuous.

In order to evaluate the contribution of fluctuations, it is necessary 
to determine the eigenfunctions of the Hamiltonian \ref{variousK}.  One 
searches for solutions to the equation 
\begin{equation}
-K_{1,2}{{d^2\theta \left( s \right)} \over {ds^2}}+F\theta \left( s \right)=
\lambda \theta \left( s \right)
\label{twoKextremum} 
\end{equation}
There will be two sorts of solution: even and odd.

\subsection*{Even solutions}

These solutions have the form
\begin{equation}
\theta(s) = \left\{ \begin{array}{ll} \cosh q_{1}s & 0 < s <l/2 \\
A\cosh q_{2}(s-l/2) + B \sinh q_{2}(s-l/2) & l/2 <s <l/2 + d \\
C \cosh q_{1}(s-l/2-d) + D \sinh q_{1}(s-l/2-d) & l/2+d<s \end{array} \right.
\label{even1}
\end{equation}	
where $\theta_{e}(-s)=\theta_{e}(s)$.

Here 
\begin{equation}
q_{1}= \sqrt{(F-\lambda)/K_{1}}
\label{q1}
\end{equation}					
and 
\begin{equation}
q_{2}= \sqrt{(F - \lambda)/K_{2}} = q_{1}\sqrt{\frac{K_{1}}{K_{2}}}	
\label{q2}
\end{equation}	
Making use of the boundary conditions, we find
\begin{eqnarray}
A & = & \cosh q_{1}l/2 \label{A} \\
B & = & \sqrt{\frac{K_{1}}{K_{2}}}\sinh q_{1}l/2 \label{B} \\
C & = & \cosh q_{1}l/2 \cosh q_{2} d + 
\sqrt{\frac{K_{1}}{K_{2}}}\sinh q_{1}l/2 \sinh q_{2} d \label{C} \\
D & = & \sqrt{\frac{K_{2}}{K_{1}}} \cosh q_{1}l/2 \sinh q_{2}d + \sinh 
q_{1}l/2 \cosh q_{2}d
\label{D}
\end{eqnarray}

\subsection*{Odd solutions}

Here, the solutions are
\begin{equation}
\theta(s) = \left\{ \begin{array}{ll} \sinh q_{1}s & 0 < s <l/2 \\
A\cosh q_{2}(s-l/2) + B \sinh q_{2}(s-l/2) & l/2 <s <l/2 + d \\
C \cosh q_{1}(s-l/2-d) + D \sinh q_{1}(s-l/2-d) & l/2+d<s \end{array} \right.
\label{odd1}
\end{equation}
where $\theta_{o}(-s) = - \theta_{o}(s)$, and, in this case, the coefficients 
are given by
\begin{eqnarray}
A & = & \sinh q_{1}l/2 \label{A1} \\
B & = & \sqrt{\frac{K_{1}}{K_{2}}}\cosh q_{1}l/2 \label{B1} \\
C & = & \sinh q_{1}l/2 \cosh q_{2} d + \sqrt{\frac{K_{1}}{K_{2}}}\cosh 
q_{1}l/2 \sinh q_{2} d \label{C1} \\
D & = & \sqrt{\frac{K_{2}}{K_{1}}} \sinh q_{1}l/2 \sinh q_{2}d + \cosh 
q_{1}l/2 \cosh q_{2}d
\label{D1}
\end{eqnarray}

The ultimate goal of the calculation is the so-called Fredholm determinant 
of the differential operator on the right hand side of Eq.  
\ref{twoKextremum}.  This determinant is readily evaluated with the use of 
a by-now well-known trick \cite{Coleman}.  The trick yields the following 
expression for the contribution to the free energy of fluctuations in 
$\theta(s)$.
\begin{equation}
A={{k_BT} \over 2}\ln \left( {\theta _e\left( L \right) \theta 
_o\left( L \right)} \right)
\label{Aresult}
\end{equation} 
where the arguments of the logarithm are the even and 
odd solutions displayed in \ref{even1} and \ref{odd1}, where 
$\lambda$ has been set equal to 0.  There are, in addition, 
normalization factors.  The factor that multiplies the even solution 
sets the magnitude of that solution equal to one at $s=0$.  The factor 
multiplying the odd solution sets the slope of that solution equal to 
unity at the origin.  The argument $L$ is the total length of the 
segment to which the proteins are attached.  In the following 
development, the limit $L \rightarrow \infty$ is taken, and the 
contribution to the free energy relevant to the interaction between 
the two attached proteins is extracted.

In the limit of asymptotically large $L$ the two factors in the argument 
of the logarithm in Eq.  \ref{Aresult} are
\begin{equation}
e^{-q_1d}\left\{ {\cosh q_2d+\sinh q_2d
\left[ {\sqrt {{{K_1} \over {K_2}}}{{1+e^{-q_1l}} \over 2}+
\sqrt {{{K_2} \over {K_1}}}{{1-e^{-q_1l}} \over 2}} \right]} \right\}
\label{fact1}
\end{equation} 
and
\begin{equation}
e^{-q_1d}\left\{ {\cosh q_2d+\sinh q_2d
\left[ {\sqrt {{{K_1} \over {K_2}}}{{1-e^{-q_1l}} \over 2}+
\sqrt {{{K_2} \over {K_1}}}{{1+e^{-q_1l}} \over 2}} \right]} \right\}
\label{fact2}
\end{equation}

After some algebra we find for the dependence of the free energy of the 
system on the parameters $l$, $d$, and the $K_{i}$'s
\begin{equation}
A={{k_BT} \over 2}\ln \left( f \right)
\label{Aexpression}
\end{equation}
where
\begin{eqnarray}
f &=& e^{2(q_{2}-q_{1})d}+\frac{e^{-2q_{1}d}\sinh 
2q_{2}d}{2}\frac{\left(\sqrt{K_{1}}-\sqrt{K_{2}}\right)^{2}}{\sqrt{K_{1}K_{2}}}
\nonumber \\ 
&&+ \frac{e^{-2q_{1}d}\sinh^{2}q_{2}d}{4} 
\frac{\left(K_{1}-K_{2}\right)^{2}}{K_{1}K_{2}}\left[1-e^{-2q_{1}l}\right]
\label{f}
\end{eqnarray}

Note that the dependence of the argument of the logarithm on the 
separation, $l$, between the two regions of differing stiffness constant 
is as
$A-Be^{-2q_1l}$.  This implies an always attractive interaction.

\end{appendix}

\pagebreak

\centerline{\Large Table 1}

\vspace{0.2in}

\centerline{\Large \bf Enthalpic energy of interaction}

\vspace{0.2in}

\begin{center}
\renewcommand{\arraystretch}{1.2}
\begin{tabular}{llll}
\hline
$F$(pN) & $\xi(F)  $ & $V_{\rm 
enth}(l=0 \ \mbox{\AA})$ & $V_{\rm enth}(l=50 \ \mbox{\AA})$ \\
 & $\xi_{p}=500 \ \mbox{\AA}$ &in kcal/mole & in kcal/mole \\
\hline \hline 
0.1 & 221 \ \AA & 0.82 & 0.66 \\
\hline 1.0 & 70 \ \AA & 2.6 & 1.27 \\
\hline 10 & 22 \ \AA & 8.2 & 0.85 \\
\hline
\end{tabular}

\end{center}

\vspace{0.2in}

\noindent Table of the enthalpic interaction strength, as given by the 
last term on the right hand side of Eq.  (\ref{E2enth}), for 
physically reasonable values of the temperature, the persistence length 
$\xi_{p}$,
and the tension, $F$.  For the dependence of the binding cooperativity 
range, $\xi(F)$ on tension and other parameters, see Eq.  (\ref{xi(F)}).

\pagebreak

\pagebreak

\begin{figure}
\caption{Strand of DNA containing two proteins.  The proteins induce 
``kinking'' on the DNA strand.  Illustrated are the case of 
antisymmetric and symmetric configurations of the two bound proteins.  
The distance between proteins is expressed in terms of the locations, 
$s_{1}$ and $s_{2}$, of each of them on the strand.}
\label{fig:twokinks}
\end{figure}
\begin{figure}
\caption{The attractive interaction for an antisymmetrically oriented 
pair of bound proteins.  Plotted is the total energy of the 
configuration in units of $k_{B}T$, multiplied by the ratio 
$\xi(F)/\xi_{p}$, where $\xi(F)$ is the tension-dependent length 
scale, defined in Eq.  \ref{xi(F)} and $\xi_{p}$ is the persistence 
length of the DNA strand.  Curves are displayed for various values of the 
``kink angle'', $\alpha$, as shown in Figure \ref{fig:twokinks}}
\label{fig:attractive}
\end{figure}
\begin{figure}
\caption{The repulsive interaction between two proteins bound in a 
symmetric configuration. Plotted is the total energy of the 
configuration in units of $k_{B}T$, multiplied by the ratio 
$\xi(F)/\xi_{p}$, where $\xi(F)$ is the tension-dependent length 
scale, defined in Eq.  \ref{xi(F)} and $\xi_{p}$ is the persistence 
length of the DNA strand. Curves are displayed for various values of the 
``kink angle'', $\alpha$, as shown in Figure \ref{fig:twokinks}}
\label{fig:repulsive}
\end{figure}
\begin{figure}
\caption{The entropic interaction potential between two identical 
bound proteins resulting from the effect of each of them on the 
bending modulus of the strand of DNA to which they are attached.  The 
interaction is plotted in units of $k_{B}T$ under the assumption that 
the region of affected DNA is 70 \AA long, and that the applied 
tension is 1 pN.}
\label{fig:Vgraph}
\end{figure}
\begin{figure}
\caption{The ``loop''  or L configuration of two symmetrically bound proteins, 
when the bend angle enforced by a bound protein is greater than 
$\pi/2$.  This is the preferred configuration when the applied tension 
is small.}
\label{fig:loop}
\end{figure}
\begin{figure}
\caption{The stretched, or S configuration of two symmetrically bound 
proteins, when the bend angle exceeds $\pi/2$.  This is the preferred 
configuration at high levels of applied tension.}
\label{fig:repulsivepair}
\end{figure}
\begin{figure}
\caption{The overall extension of a very long strand of DNA containing two 
identical symmetrically bound proteins, the bend angle, $\alpha$, of each of 
which is equal to 2.1. The extension, $X$, is relative to the fully extended DNA 
strand, measured in units of
$(2K/F)^{1/2}$, 
where $F$ is the applied tension and 
$K$ is the bending modulus. The vertical axis is the applied tension in units of 
$K/2l^2$
where $L$ is the distance along the DNA backbone between proteins 
along the DNA strand. See Appendix A for a full discussion of the parameters
utilized.}
\label{fig:transition}
\end{figure}
\begin{figure}
\caption{The parameters in Eq.  \ref{energy}.  The quantity 
$\theta(s)$ is the angle between the flexible strand and the 
horizontal axis, while $s$ is arclength.}
\label{fig:parameters}
\end{figure}
\begin{figure}
\caption{The regions of varying stiffness in a strand containing two 
bound proteins that alter the bending modulus where they attach to a 
strand of DNA.  The values that the bending modulus, $K$, takes in the 
various regions are indicated.  See Eq.  \ref{variousK} and 
surrounding text in Appendix B.}
\label{fig:stiffness}
\end{figure}

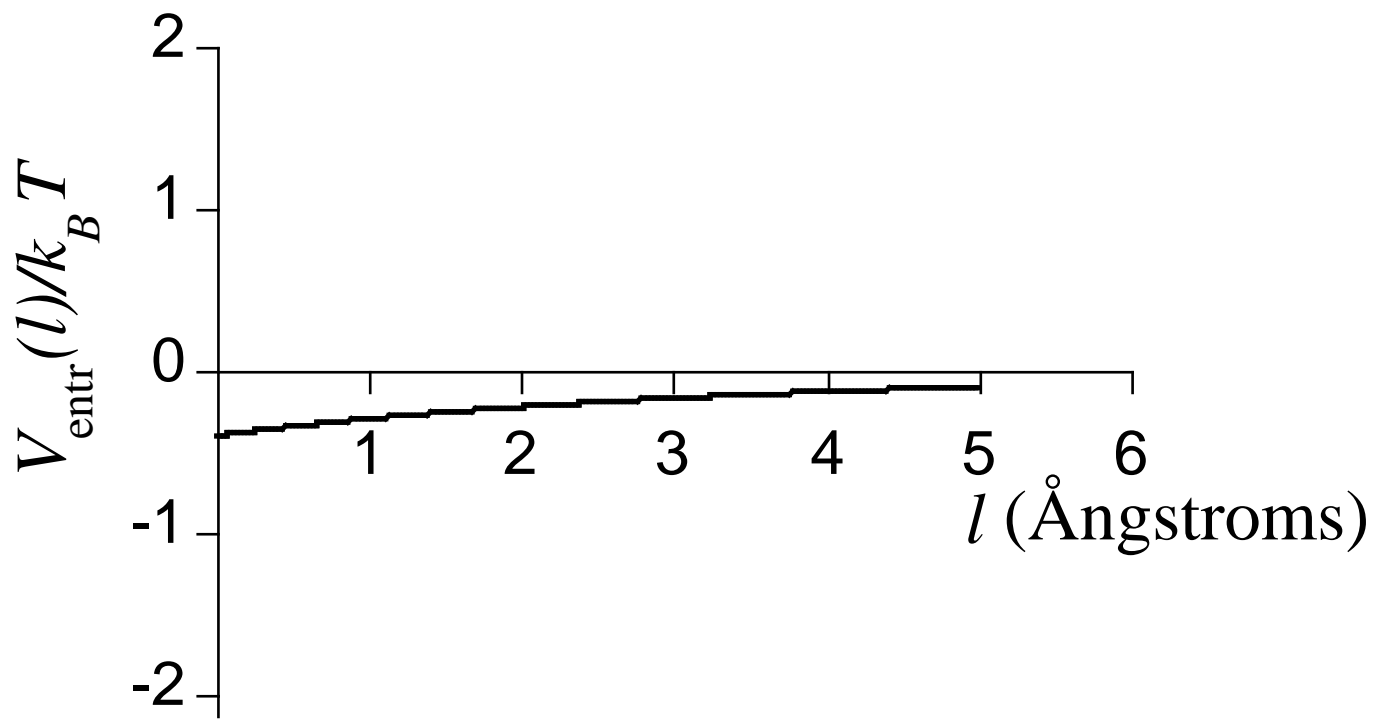
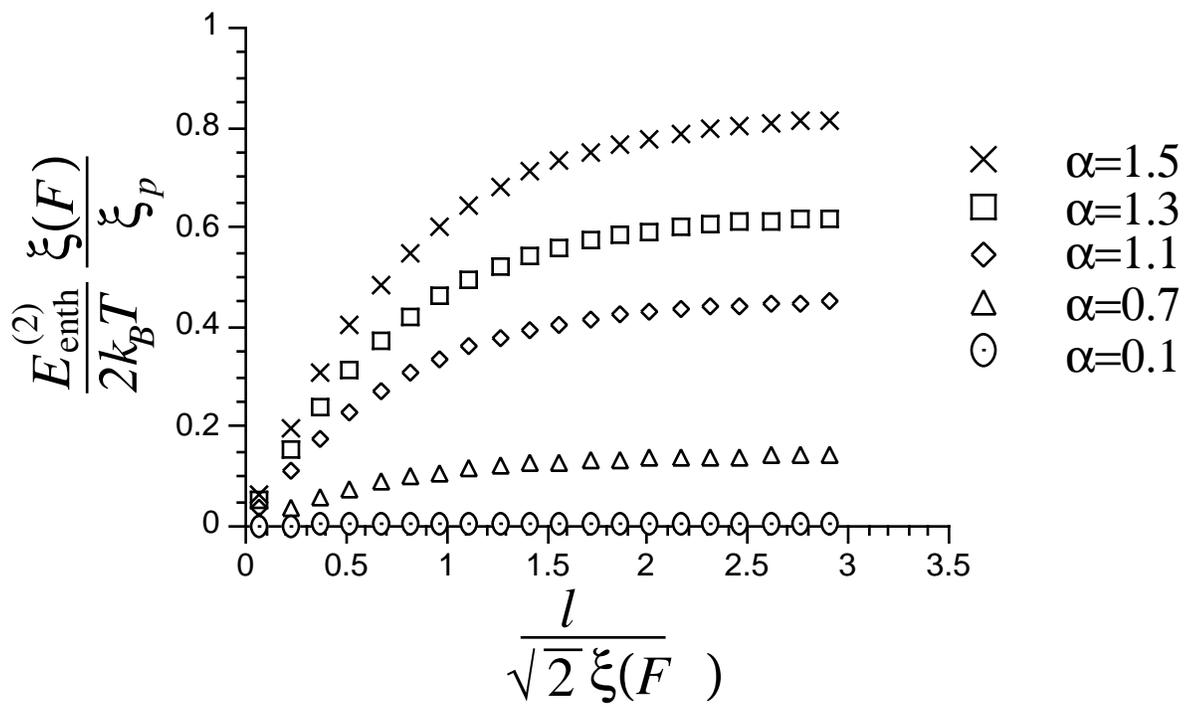
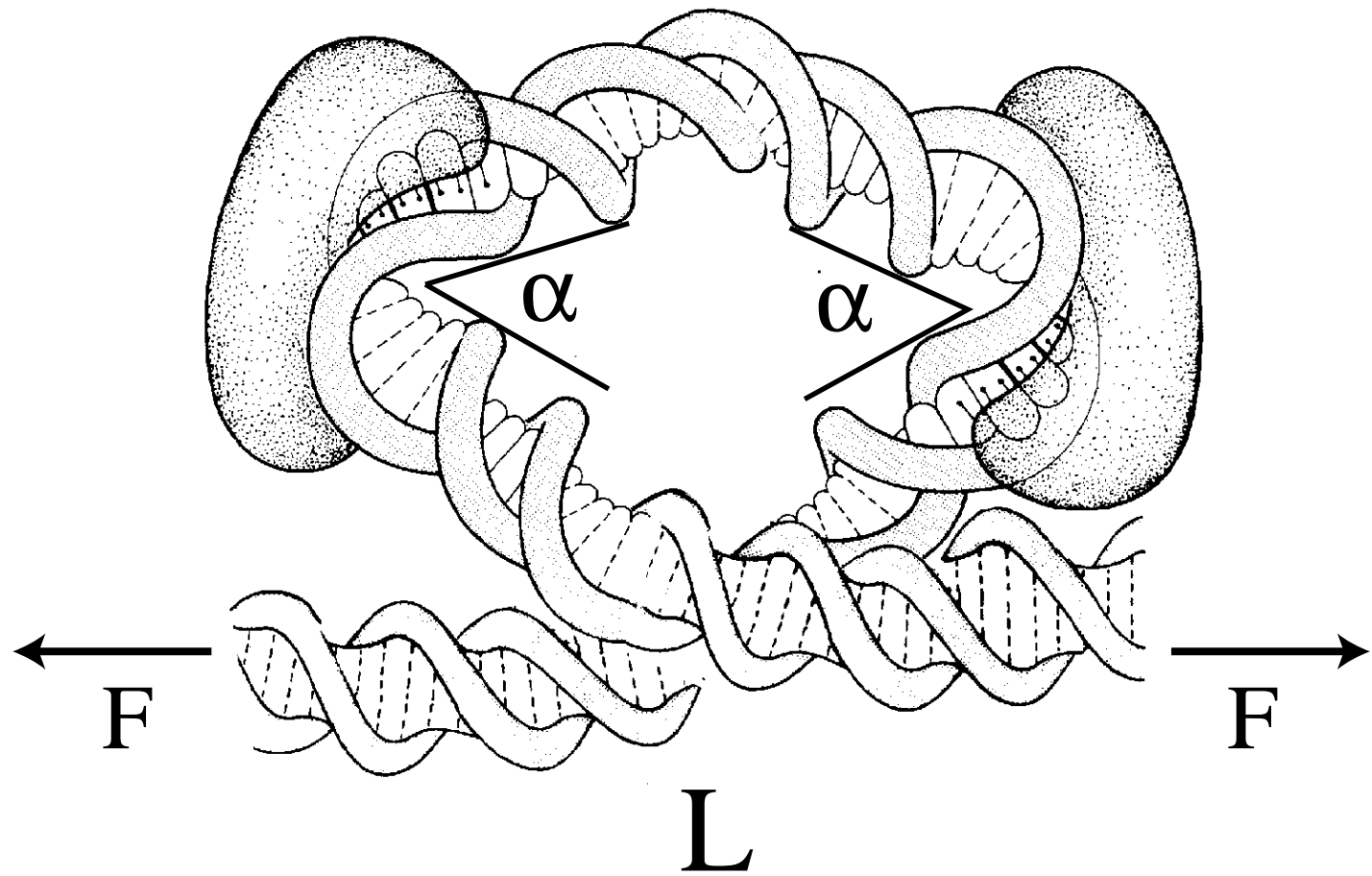
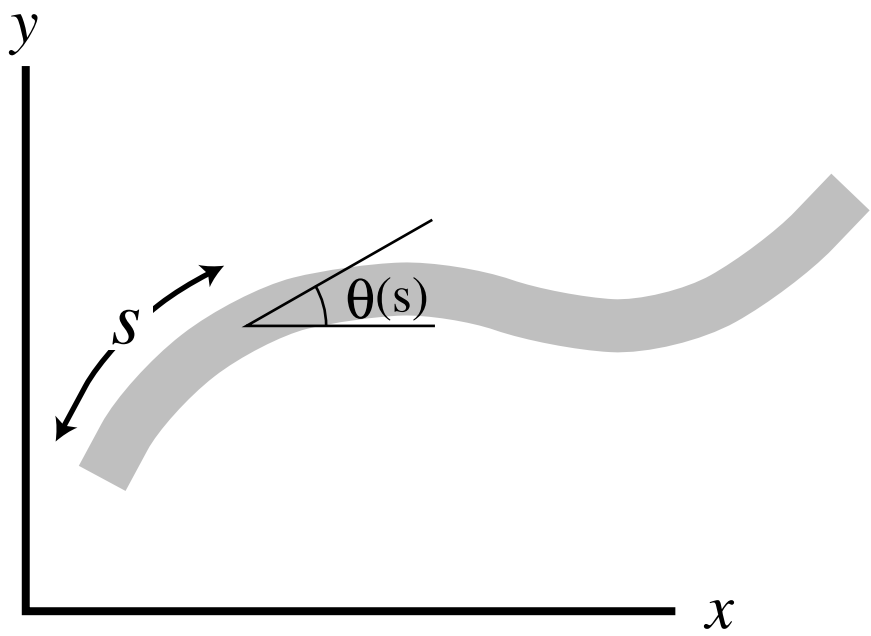
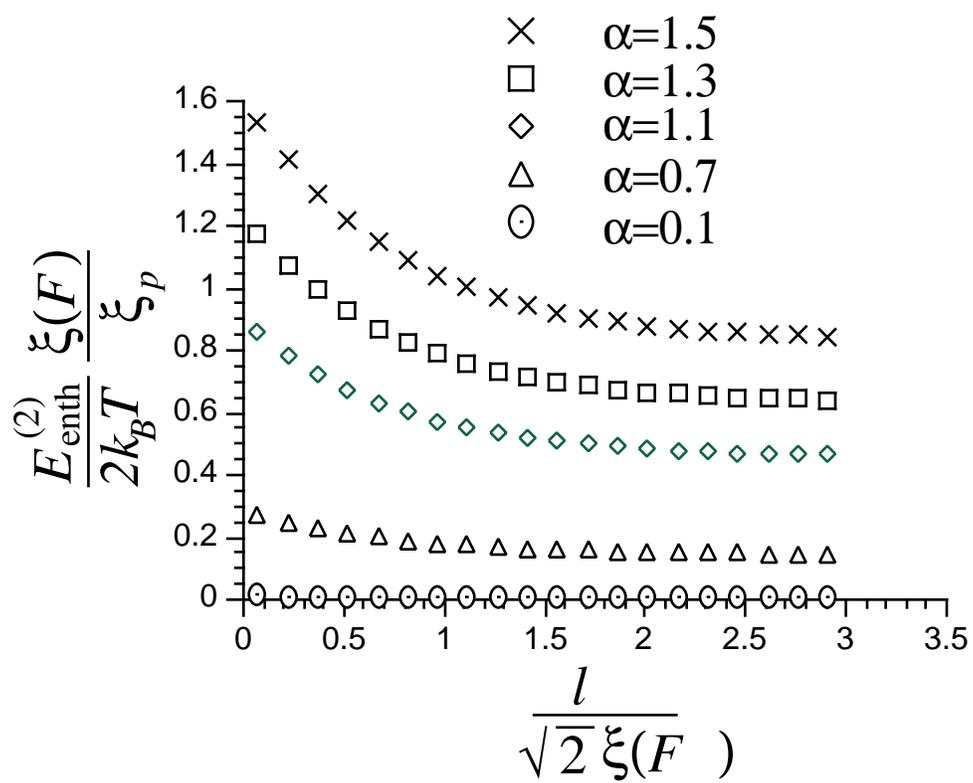
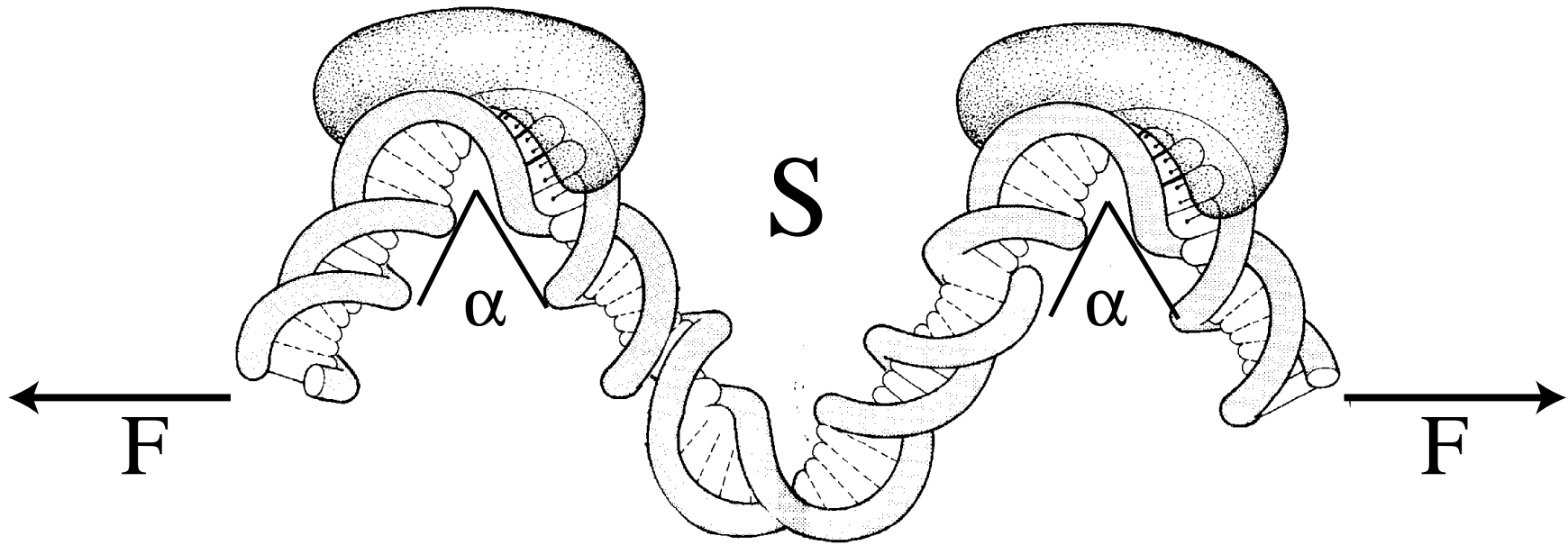
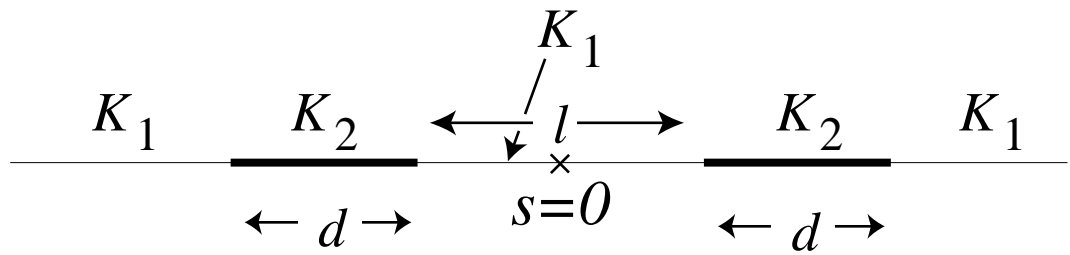
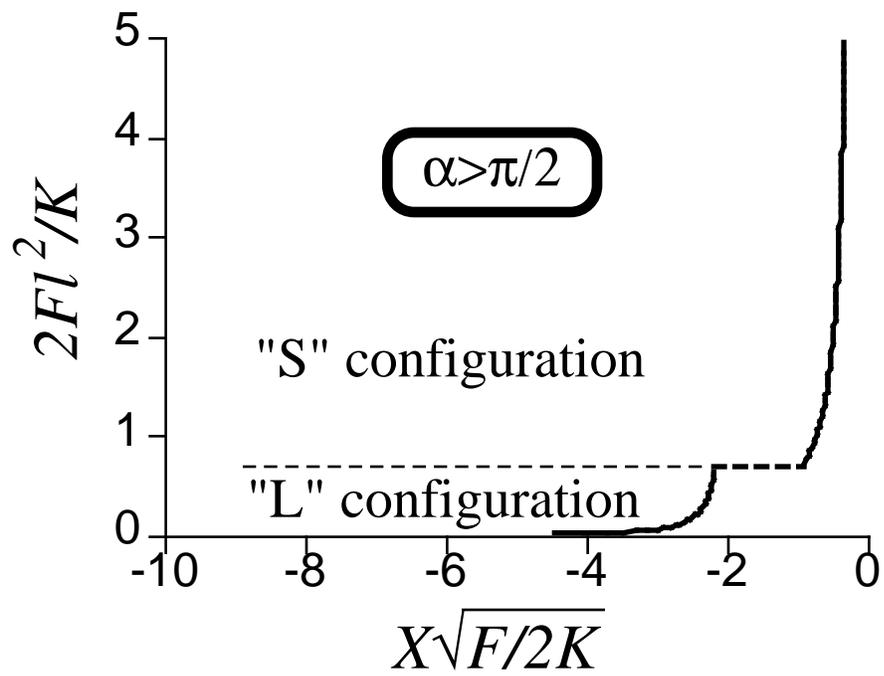
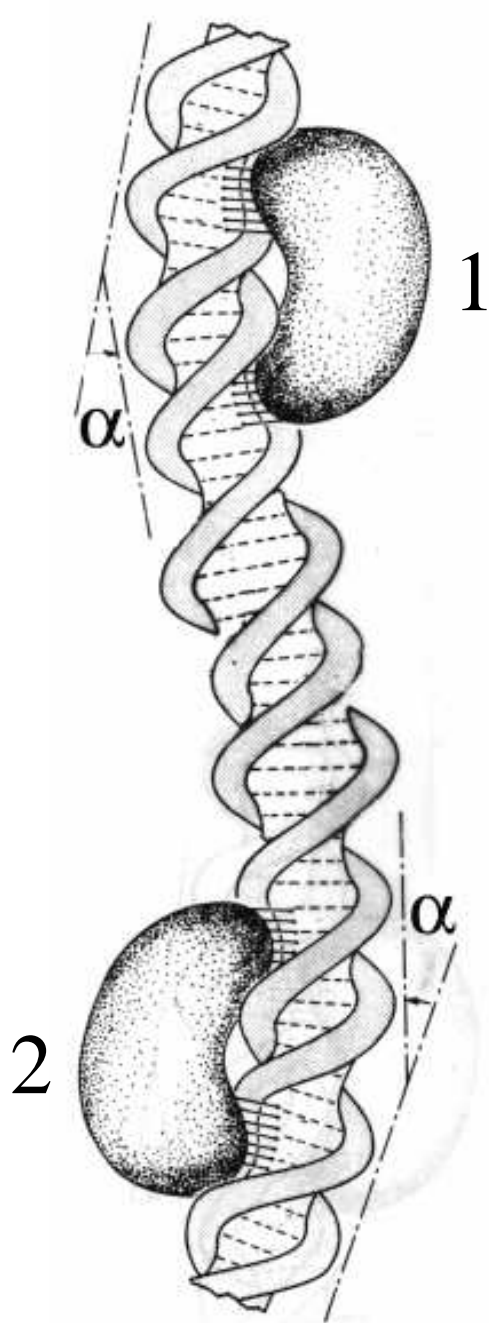
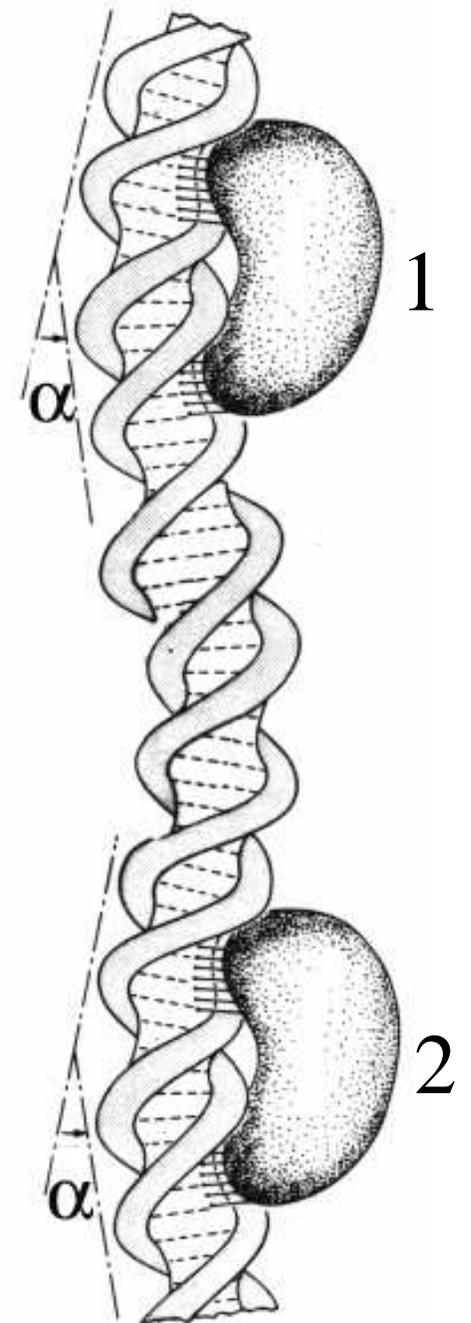
\end{document}